\documentclass[11pt]{article}
\usepackage{amsmath}
\usepackage{amssymb}
\usepackage{amsthm}
\usepackage{enumerate}
\usepackage{bbm}
\usepackage{algorithm}
\usepackage{algorithmic}
\usepackage{comment}
\usepackage{hyperref}
\usepackage{cleveref}
\usepackage[round]{natbib}
\usepackage{color}
\usepackage[margin=1in]{geometry}
\usepackage{multirow}
\usepackage{commath}
\usepackage{nicefrac}

\newtheorem{theorem}{Theorem}[section]
\newtheorem{lemma}{Lemma}[section]

\newtheorem{example}{Example}[section]

\theoremstyle{remark}
\newtheorem*{remark}{Remark}

\newcommand{\xx}{{\mathbf{x}}}

\newcommand{\pp}{{\mathbf{p}}}

\newcommand{\Real}{{\ensuremath{\mathbb{R}}}}

\newcommand{\sw}{{\mathrm{SW}}}
\newcommand{\profit}{{\mathrm{Profit}}}

\DeclareMathOperator*{\argmax}{arg\,max}

\author{Ziwei Ji\thanks{University of Illinois at Urbana-Champaign. ziweiji2@illinois.edu}
\and Ruta Mehta\thanks{University of Illinois at Urbana-Champaign. rutameht@illinois.edu}
\and Matus Telgarsky\thanks{University of Illinois at Urbana-Champaign. mjt@illinois.edu}}
\title{Social welfare and profit maximization from revealed preferences}
\date{}
\begin{document}

\maketitle

\begin{abstract}
Consider the {\em seller's problem} of finding optimal prices for her $n$ (divisible) goods when faced with a set of $m$ consumers, given that she can only observe their purchased bundles at posted prices, i.e., \emph{revealed preferences}. We study both social welfare and profit maximization with revealed preferences.
Although social welfare maximization is a seemingly non-convex optimization problem in prices, we show that $(i)$ it can be reduced to a dual convex optimization problem in prices, and $(ii)$ the revealed preferences can be interpreted as supergradients of the concave conjugate of valuation, with which subgradients of the dual function can be computed. We thereby obtain a simple subgradient-based algorithm for strongly concave valuations and convex cost, with query complexity $O(m^2/\epsilon^2)$, where $\epsilon$ is the additive difference between the social welfare induced by our algorithm and the optimum social welfare.
We also study social welfare maximization under the online setting, specifically the random permutation model, where consumers arrive one-by-one in a \emph{random order}. For the case where consumer valuations can be arbitrary continuous functions, we propose a price posting mechanism that achieves an expected social welfare up to an additive factor of $O(\sqrt{mn})$ from the maximum social welfare.
Finally, for profit maximization (which may be non-convex in simple cases), we give nearly matching upper and lower bounds on the query complexity for separable valuations and cost (i.e., each good can be treated independently).
\end{abstract}

\section{Introduction}

In consumer theory, it is standard to assume that the preferences of a consumer are captured by a \emph{valuation} function, which is often assumed to be known to the mechanism designer. However, in a real market, one can only observe what buyers buy at given prices, the \emph{revealed preferences}. Research on revealed preferences within TCS has two primary objectives: $(i)$ learning valuation functions from revealed preferences, with the goal of having predictive properties \citep{BV06,ZR12,BDMUV14,BMM15}; $(ii)$ directly learning the prices that maximizeѕ social welfare or profit \citep{dB15,RUW16,RSUW17,BDKS15,BZ09,BZ12,BR12,WDY14,ACDKR15}.

The latter problem is of importance to sellers in today's online economies, where a large amount of data about consumers' buying patterns is available. For a seller, profit maximization is the primary goal in general, while she may also want to maximize social welfare in an effort to earn the goodwill of consumers, with increased market share as a byproduct.

In this paper, we consider social welfare and profit maximization using only revealed preferences.

\subsection{Our model, results, and techniques}

Consider a market with $m$ consumers and a producer (seller) who produces and sells a set of $n$ \emph{divisible} goods. In the most general case, the preferences of consumer $i$ over bundles of goods are defined by a valuation function $v_i:\mathcal{C} \rightarrow \Real_+$ ($\mathcal{C}\subset \mathbb{R}^n$ is called the feasible set), which is her private information and \emph{unknown}. At prices $\pp$ she demands bundle $\xx_i(\pp)$ that maximizes her value minus payment, i.e., her {\em quasilinear utility}
\begin{equation*}
\xx_i(\pp) \in \argmax_{\mathbf{x}\in \mathcal{C}}\del{v_i(\xx) - \langle \pp, \xx\rangle}.
\end{equation*}
Given prices $\pp$, the \emph{revealed preference} refers to the purchased bundle $\xx_i(\pp)$ of each consumer in the market (\emph{demand oracle information}), or even only $\sum_{i=1}^{m}\mathbf{x}_i(\mathbf{p})$ (\emph{aggregate demand oracle information}). No other information of the valuations is revealed.

Producing the demanded goods incurs cost to the producer, which is represented by a convex cost function $c$. The producer, or the algorithm, posts prices and makes observations repeatedly, trying to maximize the \emph{social welfare} or \emph{profit}, as described below.

\paragraph{Social welfare maximization.}

The social welfare of bundles $\xx_1,\ldots,\xx_m\in \mathcal{C}$ is the sum of consumers' valuations minus production cost, i.e.,
\begin{equation}\label{socWelf}
    \sw(\xx_1,\ldots,\xx_m)=\textstyle\sum_{i=1}^{m}v_i(\xx_i)-c\left(\textstyle\sum_{i=1}^{m}\xx_i\right).
\end{equation}
The benchmark used in this paper is the maximum social welfare $\mathrm{SW}^*$ and corresponding maximizing bundles $(\mathbf{x}_1^*, \ldots,\mathbf{x}_m^*)$, defined as
\begin{equation}\label{socWelfMax}
    \mathrm{SW}^*=\max_{\mathbf{x}_1,\ldots,\mathbf{x}_m\in \mathcal{C}}\mathrm{SW}(\mathbf{x}_1,\ldots,\mathbf{x}_m)\quad\mathrm{and}\quad(\mathbf{x}_1^*,\ldots,\mathbf{x}_m^*)\in\argmax_{\mathbf{x}_1,\ldots,\mathbf{x}_m\in \mathcal{C}}\sw(\mathbf{x}_1,\ldots,\mathbf{x}_m).
\end{equation}

In \Cref{sec:offlineSW}, offline social welfare maximization is considered. The producer tries to find good prices $\mathbf{p}$ such that $\sw(\mathbf{x}_1(\mathbf{p}),\ldots,\mathbf{x}_m(\mathbf{p}))$ is maximized. Although there exist many methods to maximize a concave function, the social welfare is usually a non-concave function in $\mathbf{p}$ \citep{RUW16}. Moreover, the producer only has access to the aggregate demand oracle; the true valuations $v_i(\mathbf{x}_i(\mathbf{p}))$ are unknown.

We first show using duality theory that the maximum social welfare $\sw^*$, which is larger than or equal to any social welfare that can be induced by some prices, can in fact be induced by a single price vector $\mathbf{p}^*$, which is the minimizer of a convex dual function $f(\mathbf{p})=c^*(\mathbf{p})-\sum_{i=1}^{m}v_i^*(\mathbf{p})$, where $c^*$ and $v_i^*$ are respectively {\em convex} and {\em concave conjugates} \citep{ROC}, as reviewed in \Cref{sec:common}. Moreover, the revealed preferences are supergradients of $v_i^*$, with which subgradients of $f$ can be computed. Finally, to get a faster algorithm, we apply a smoothing technique to $f$ and then invoke the accelerated gradient descent method. These ideas are formalized in \Cref{offlineSWMaxAlg}, whose guarantee is given below.
\begin{theorem}[Informal statement of \Cref{offlineGuarantee}]\label{offlineGuaranteeIntro}
    The additive error between the social welfare induced by \Cref{offlineSWMaxAlg} and the maximum social welfare \cref{socWelfMax} is at most $O(m/\sqrt{T})$, where $T$ is the number of queries to the aggregate demand oracle.
\end{theorem}
In other words, to ensure an additive $\epsilon$ approximation of the maximum social welfare, \Cref{offlineSWMaxAlg} needs $O(m^2/\epsilon^2)$ queries to the aggregate demand oracle.

\citep{RUW16} and \citep{RSUW17} are the most relevant prior work. \citep{RUW16} studies profit maximization instead of social welfare maximization in a market with one consumer. However, it is assumed that the valuation function is homogeneous, under which profit maximization can be reduced to social welfare maximization. Assumptions made in \citep{RUW16} and this paper are basically identical. Key differences are: (i) \citep{RUW16} proposes a two-level algorithm, where there is an outer iterative algorithm maximizing social welfare, and for each outer iterate, the supergradient of the unknown valuation function is computed by solving a dual optimization problem. In this paper, we only need to solve a single (different) dual optimization problem. Therefore, this gives a simpler approach which may be of independent interest. (ii) The subgradient of the dual objective function in this paper can be interpreted as the excess supply (see the discussion around \cref{eq:RPSubgrad} in \Cref{sec:offlineSWAlg}), which gives our algorithm a natural interpretation as a T\^{a}tonnement process. (iii) The query complexity given in \citep{RUW16} can be as large as $O(1/\epsilon^6)$ to ensure an additive error of $\epsilon$ between the induced social welfare and the maximum social welfare; one reason is that they use subgradient descent, which works for non-smooth convex functions but converges slowly. In this paper, by combining a smoothing technique and accelerated gradient descent, \Cref{offlineGuaranteeIntro} only needs $O(1/\epsilon^2)$ queries to the aggregate demand oracle. \citep{RSUW17} assumes that the valuation is stochastic, but only considers a linear cost. It also considers unit demand consumers with indivisible goods, which is out of the scope of this paper.


Next in \Cref{sec:onlineSW} we consider \emph{online} social welfare maximization under the \emph{random permutation} model. In this model, $m$ consumers come to make purchases one by one, and correspondingly the producer is allowed to post prices dynamically, i.e., to update prices from $\mathbf{p}_i$ to $\mathbf{p}_{i+1}$ after the purchase of consumer $i$. Random permutation here means that those $m$ consumers are first chosen potentially by an adversary, and then come and make purchases one by one in a uniformly random order. (The random permutation model has been extensively studied within online optimization \citep{GM08,DH09,AD15}, and is more general than the i.i.d. model where each valuation is an independent sample from an unknown distribution.) The objective is to maximize the expected online social welfare $\mathbb{E}[\sw(\mathbf{x}_1(\mathbf{p}_1),\ldots,\mathbf{x}_m(\mathbf{p}_m))]$, where the expectation is taken over random orders.

The idea to solve the online social welfare maximization problem is to run an online convex optimization algorithm on a dual problem $f_i(\mathbf{p})=\nicefrac{c^*(\mathbf{p})}{m}-\langle \mathbf{x}_i(\mathbf{p}_i),\mathbf{p}\rangle$. See \Cref{onlineSWMaxAlg} for details; an introduction to online convex optimization is given in \Cref{sec:onlineSW}.
\begin{theorem}[Informal statement of \Cref{onlineGuarantee}]
    The expected additive error between the online social welfare achieved by \Cref{onlineSWMaxAlg} and the maximum offline social welfare \cref{socWelfMax} is bounded by $O(\sqrt{mn})$, where the expectation is taken over random orders of valuations.
\end{theorem}
For a given producer, the number of goods $n$ can be thought of as fixed. As a result, the loss of social welfare induced by \Cref{onlineSWMaxAlg} is sublinear in the number of consumers $m$.

The idea of \Cref{onlineSWMaxAlg} comes from \citep{AD15}, where a general online stochastic convex programming problem is considered. It has many other advantages when applied to online social welfare maximization. First, it is enough to assume that the valuations are continuous; the consumer demand oracle may potentially need to solve some non-convex quasi-linear utility maximization problem, but our focus is on the producer side. Since $f_i$ only depends on the revealed preference $\mathbf{x}_i(\mathbf{p}_i)$, not on $v_i$, it is still convex. Second, \Cref{onlineSWMaxAlg} is robust, in the sense that it is not sensitive to the potential error in quasi-linear utility maximization. For details, see the discussion at the end of \Cref{sec:onlineSW}.

\paragraph{Profit maximization.}

Next we consider profit maximization with access to the aggregate demand oracle. Given prices $\mathbf{p}$, the profit of producer is the revenue minus production cost, i.e.,
\begin{equation}\label{eq:profit}
    \profit(\pp) = \langle \pp, \textstyle\sum_{i=1}^m \xx_i(\pp)\rangle - c\left(\textstyle\sum_{i=1}^m \xx_i(\pp)\right).
\end{equation}

Although it is more reasonable for the producer to maximize the profit, this problem is hard due to non-convexity. The social welfare maximization problems are solved by making a reduction to some convex optimization problem on the space of prices. However, for profit maximization, both the set of optimal bundles and the set of optimal prices may be non-convex, as shown by \Cref{eg} in \Cref{sec:profit}.

We then consider the case where both the valuations and cost are separable. A separable valuation  $v_i(\mathbf{x})=\sum_{j=1}^{n}v_{ij}(x_j)$, while similarly a separable cost $c(\mathbf{y})=\sum_{j=1}^{n}c_j(y_j)$. Under this assumption, in \Cref{sec:profit} we give upper and lower bounds on the query complexity for profit maximization and revenue maximization (i.e., the cost is $0$). These upper and lower bounds match for revenue maximization.

\begin{theorem}[Informal statement of \Cref{thm:profitUpper,thm:profitLower}]
    Consider a market with $m$ consumers and $n$ goods. If the valuations are strongly concave, and
    both the valuations and cost are separable and Lipschitz continuous, then Algorithm \ref{profitMaxAlg} maximizes the profit up to an additive $\epsilon$ error with $O(mn/\epsilon)$ queries to the aggregate demand oracle. If the cost is zero, then the strongly concave assumption on valuations can be dropped.

    On the other hand, for concave, separable and Lipschitz continuous valuations, any algorithm requires $\Omega(n/\epsilon)$ queries to the aggregate demand oracle in order to maximize the revenue up to an $\epsilon$ additive error.
\end{theorem}

\subsection{Related work}

Samuelson started the theory of {\em revealed preferences} in 1938 \citep{S38} to facilitate mapping observed data to valuation functions, which led to extensive work within economics on ``rationalization'' or ``fitting the samples'' \citep{HSH,Rich,Afriat,uzawa, MC-RP1,MC-RP2,die,varian05}. In TCS, there have been a lot of work on learning valuations from revealed preferences with which predictions can be made \citep{BV06,ZR12,BDMUV14,BMM15}.

Another line of research is on learning prices directly that can maximize social welfare or profit, usually known as the dynamic pricing problem \citep{dB15,BDKS15,BZ09,BZ12,BR12,ACDKR15}.
Some prior works assume nice properties of the demand function (oracle) itself, such as linearity in case of large number of goods \citep{KZ14,dBZ13}, concavity \citep{BDKS15}, Lipschitz continuity \citep{BZ09,BZ12,WDY14}. However, these properties may not be satisfied by demands that come from typical concave valuation functions. In \citep{ACDKR15}, the valuation function is assumed to be linear, and is first partially inferred and then used in a price optimization step. However, if the valuation is general concave, such a learning phase is not possible \citep{BV06}. 

Recently, \citep{DRSWW18} studies an online linear classification problem under the revealed preference model.

\section{Preliminaries}\label{sec:common}

\paragraph{Market model.}

Our model consists of one producer (seller) who produces and sells $n$ divisible goods, and $m$ consumers. Consumer $i$'s preferences are represented by an \emph{unknown} valuation function $v_i:\mathcal{C}\to \Real_+$. The feasible consumption set $\mathcal{C}\subset \mathbb{R}_+^n$ is typically assumed to be convex and compact with non-empty interior. It is assumed that $\mathcal{C}$ is known to the algorithm, and let $D=\max_{\mathbf{x},\mathbf{y}\in \mathcal{C}}\|\mathbf{x}-\mathbf{y}\|_2$ denote the $\ell_2$ diameter of $\mathcal{C}$. Note that our algorithm can be extended to the case where $v_i$'s have different domains with different diameters; a common domain $\mathcal{C}$ is used here only for convenience.

Given prices $\pp=(p_1,\dots,p_n) \in \Real_+^n$ of goods, the quasi-linear utility of a bundle $\mathbf{x}\in \mathbb{R}_+^n$ is defined as
\[
u_i(\mathbf{x},\mathbf{p})=v_i(\mathbf{x})-\langle \mathbf{x},\mathbf{p}\rangle.
\]
Naturally, consumer $i$ demands a bundle from $\mathcal{C}$ that maximizes her quasi-linear utility
\[
\xx_i(\pp) \in \argmax_{\xx \in \mathcal{C}} u_i(\xx,\pp),
\]
which is known as the \emph{revealed preference} of consumer $i$ at prices $\pp$. Once the seller sets prices $\pp$, we only get to see $\xx_i(\pp)$ for each consumer $i$, where every consumer can be thought of as a \emph{demand oracle}, or even only $\mathbf{x}(\mathbf{p})=\sum_{i=1}^{m}\mathbf{x}_i(\mathbf{p})$, where the market can be seen as an \emph{aggregate demand oracle}. $v_i$ is always assumed to be continuous to ensure that $\mathbf{x}_i(\mathbf{p})$ exists. This is the only assumption needed for the online social welfare maximization part of this paper; the offline social welfare maximization part and the profit maximization part further assume that the valuations are strongly concave, which will be introduced later.

The production cost is represented by a convex, Lipschitz continuous, non-decreasing cost function $c:m\mathcal{C}\to\Real_+$, where
$m\mathcal{C} =\{m \mathbf{x}|\mathbf{x}\in \mathcal{C}\}=\{\sum_{i=1}^{m}\mathbf{x}_i|\mathbf{x}_1,\ldots,\mathbf{x}_m\!\in\!\mathcal{C}\}$ since $\mathcal{C}$ is convex. Note that the domain of $c$ is big enough to allow production of any aggregate demand. Let $\lambda$ denote the modulus of Lipschitz continuity of $c$ with respect to the $\ell_2$-norm. It is assumed that the cost function is known to the algorithm.

The producer, or the algorithm, can only post prices and observe the purchased bundles repeatedly, trying to maximize the \emph{social welfare} \cref{socWelf} or \emph{profit} \cref{eq:profit}. Note that if valuations are only continuous, $\mathbf{x}_i(\mathbf{p})$ and the induced social welfare and profit may not be uniquely defined. In this paper, the online social welfare result holds for any $\mathbf{x}_i(\mathbf{p})$, while in offline social welfare and profit maximization, $\mathbf{x}_i(\mathbf{p})$ is unique since strong concavity is assumed.

\paragraph{Convex and concave conjugates.}
The notion of convex and concave conjugates are crucial in our algorithms. Given a convex function $f:\mathcal{D}\to \mathbb{R}$ where $\mathcal{D}\subset \mathbb{R}^n$ is non-empty, its convex conjugate $f^*$ is defined as:
\begin{equation*}
    f^*(\mathbf{y})=\sup_{\mathbf{x}\in \mathcal{D}}\del{\langle \mathbf{y},\mathbf{x}\rangle-f(\mathbf{x})},
\end{equation*}
where the domain of $f^*$ is given by $\mathbf{dom}\,f^*=\{\mathbf{y}\in \mathbb{R}^n|f^*(\mathbf{y})<\infty\}$. Similarly, given a concave function $f:\mathcal{D}\to \mathbb{R}$, its concave conjugate is defined as
\begin{equation*}
    f^*(\mathbf{y})=\inf_{\mathbf{x}\in \mathcal{D}}\del{\langle \mathbf{y},\mathbf{x}\rangle-f(\mathbf{x})},
\end{equation*}
where the domain of $f^*$ is given by $\mathbf{dom}\,f^*=\{\mathbf{y}\in \mathbb{R}^n|f^*(\mathbf{y})>-\infty\}$. Since we only compute convex conjugates of convex functions and concave conjugates of concave functions, the above notation is fine. $\partial f$ denotes the set of subgradients of convex $f$ or supergradients of concave $f$.

Note that in our case, since $\mathbf{dom}\,c=m\mathcal{C}$ is non-empty and compact and $c$ is continuous, $\mathbf{dom}\,c^*=\mathbb{R}^n$. Similarly, for every $i$, $\mathbf{dom}\,v_i^*=\mathbb{R}^n$.

\Cref{conjgtSubgrad} is crucial in our algorithm: One key observation in this paper is that revealed preferences are actually supergradients of the concave conjugate of valuation, which is given by \Cref{conjgtSubgrad}. \Cref{conjgtSubgrad} can be derived from \citep{HL12} Corollary E.1.4.4 immediately. Although it is stated for convex functions and convex conjugates, corresponding properties hold for concave functions and concave conjugates.
\begin{lemma}\label{conjgtSubgrad}
    Suppose $f$ is convex continuous with non-empty domain. For every pair $(\mathbf{x},\mathbf{y}) \in \mathbf{dom}\,f\times \mathbf{dom}\,f^*$,
    \begin{equation*}
        \renewcommand{\arraystretch}{1.5}
        \begin{array}{cl}
             & \mathbf{y} \in \partial f(\mathbf{x}) \\
            \iff & \mathbf{x} \in \partial f^*(\mathbf{y}) \\
            \iff & \mathbf{x} \in \arg\max_{\mathbf{x}'\in \mathbf{dom}\,f} \del{\langle \mathbf{y},\mathbf{x}'\rangle - f(\mathbf{x}') } \\
            \iff & \mathbf{y} \in \arg\max_{\mathbf{y}'\in \mathbf{dom}\,f^*} \del{ \langle \mathbf{x},\mathbf{y}'\rangle - f^*(\mathbf{y}') }\\
            \iff & f(\mathbf{x})+f^*(\mathbf{y})=\langle \mathbf{x},\mathbf{y}\rangle.
        \end{array}
    \end{equation*}
\end{lemma}

\section{Offline social welfare maximization}\label{sec:offlineSW}

\paragraph{Problem description.}

The goal of offline social welfare maximization is to find prices $\mathbf{p}\in \mathbb{R}_+^n$ such that the induced social welfare $\mathrm{SW}(\mathbf{x}_1(\mathbf{p}),\ldots,\mathbf{x}_m(\mathbf{p}))=\sum_{i=1}^{m}v_i(\mathbf{x}_i(\mathbf{p}))-c(\sum_{i=1}^{m}\mathbf{x}_i(\mathbf{p}))$ is maximized. As introduced below, $v_i$'s are assumed to be strongly concave, and thus $\textbf{x}_i(\textbf{p})$'s are uniquely determined.

\paragraph{Strongly concave valuations.}

In the offline setting, the valuation functions ($v_i$'s) are further assumed to be $\alpha$-strongly concave, meaning that $v_i(x)+\nicefrac{\alpha\|x\|_2^2}{2}$ is concave. Concavity is a standard assumption on valuations to capture diminishing marginal returns. Strong concavity, as its name suggests, is a strong assumption; however it is satisfied by many common valuations such as the constant elasticity of substitution functions and Cobb-Douglas functions (c.f. \citep{RUW16}). Furthermore, a common modulus of strong concavity $\alpha$ is only for convenience; the algorithm can be easily adapted to the case where different $v_i$ have different moduli of strong concavity.

The dual notion to strong concavity (convexity) is strong smoothness. $f:\mathcal{D}\to \mathbb{R}$ is $\beta$-strongly smooth if $f$ is differentiable and its gradient is $\beta$-Lipschitz continuous, or formally, for any $\mathbf{x},\mathbf{y}\in \mathcal{D}$, $\|\nabla f(\mathbf{y})-\nabla f(\mathbf{x})\|_2\le\beta\|\mathbf{y}-\mathbf{x}\|_2$.

The following lemma can be immediately derived from \citep{HL12} Theorem E.4.2.1 and E.4.2.2.
\begin{lemma}\label{strongConvSmoothDual}
    Suppose $f:\mathcal{D}\to \mathbb{R}$ is concave continuous and $\mathcal{D}\subset \mathbb{R}^n$ is non-empty. Then $f$ is $\alpha$-strongly concave if and only if $f^*$ is $\nicefrac{1}{\alpha}$-strongly smooth and concave on $\mathbb{R}^n$.
\end{lemma}

\paragraph{Accelerated gradient descent.}

The accelerated gradient descent algorithm, which was first introduced in \citep{N83}, gives the optimal convergence rate for smooth convex optimization problems. There have been many extensions to AGD, including \citep{T08,AO14}. In this paper, one variant called the AGM algorithm given in \citep{AO14} will be invoked.
\begin{lemma}[\citep{AO14} Theorem 4.1]\label{AGM}
    Suppose $f:\mathcal{D}\to \mathbb{R}$ is $\beta$-strongly smooth and convex and $\mathbf{x}^*\in\arg\min_{\mathbf{x}\in \mathcal{D}}f(\mathbf{x})$. Given $\mathbf{x}_0\in \mathcal{D}$, for any $t\ge1$, The AGM algorithm outputs $\mathbf{x}_t\in \mathcal{D}$ such that
    \begin{equation*}
        f(\mathbf{x}_t)-f(\mathbf{x}^*)\le \frac{2\beta\|\mathbf{x}_0-\mathbf{x}^*\|_2^2}{t^2}.
    \end{equation*}
\end{lemma}

\subsection{Algorithm and analysis}\label{sec:offlineSWAlg}

We propose \Cref{offlineSWMaxAlg} to solve the offline social welfare maximization problem.
\begin{algorithm}
    \caption{Offline SW Maximization}
    \begin{algorithmic}
        \STATE{$\mu\leftarrow\frac{2\lambda}{mD\sqrt{T}}$, $c_{\mu}(\mathbf{y})=c(\mathbf{y})+\frac{\mu}{2}\|\mathbf{y}\|_2^2$, $f_{\mu}=c_{\mu}^*(\mathbf{p})-\sum_{i=1}^{m}v_i^*(\mathbf{p})$.}
        \STATE{$\mathcal{P}\leftarrow\{\mathbf{p}\ge \mathbf{0}|\|\mathbf{p}\|_2\le\lambda\}$, $\mathbf{p}_{\mu}^0=\mathbf{0}$.}
        \STATE{Give $T$, the total number of rounds, and $\mathbf{p}_{\mu}^0$ to the AGM algorithm.}
        \STATE{Output $\mathbf{p}_{\mu}^T$ returned by the AGM algorithm.}
    \end{algorithmic}
    \label{offlineSWMaxAlg}
\end{algorithm}
\begin{theorem}\label{offlineGuarantee}
    The social welfare induced by $\mathbf{p}_{\mu}^T$ given by \Cref{offlineSWMaxAlg} is within $\nicefrac{9\lambda mD}{\sqrt{T}}+\nicefrac{16\lambda^2m}{\alpha T}$ from the maximum offline social welfare.
\end{theorem}

Here is a proof sketch; detailed proofs are given in \Cref{app_sec:offlineSW}. The proof is based on two observations. The first one, formalized in \Cref{lem:SWDuality}, says that the optimal solution of a dual optimization problem can induce the maximum social welfare $\mathrm{SW}^*$.
\begin{lemma}\label{lem:SWDuality}
    Given concave continuous valuations $v_1,\ldots,v_m:\mathcal{C}\to \mathbb{R}$ and a convex continuous cost $c:m\mathcal{C}\to \mathbb{R}$,
    \begin{equation*}
        \mathrm{SW}^*=\min_{\mathbf{p}\in \mathbb{R}^n}\del{c^*(\mathbf{p})-\sum_{i=1}^{m}v_i^*(\mathbf{p})},
    \end{equation*}
    and for any dual optimal solution $\mathbf{p}^*$, $(\mathbf{x}_1(\mathbf{p}^*),\ldots,\mathbf{x}_m(\mathbf{p}^*))$ maximizes social welfare. If furthermore the cost is non-decreasing and $\lambda$-Lipschitz continuous, then
    \begin{equation*}
        \mathrm{SW}^*=\min_{\mathbf{p}\in \mathbb{R}^n}\left(c^*(\mathbf{p})-\sum_{i=1}^{m}v_i^*(\mathbf{p})\right)=\min_{\mathbf{p}\ge \mathbf{0},\|\mathbf{p}\|_2\le\lambda}\left(c^*(\mathbf{p})-\sum_{i=1}^{m}v_i^*(\mathbf{p})\right),
    \end{equation*}
    and for any optimal solution $\mathbf{p}^*$ of the rightmost dual problem, $(\mathbf{x}_1(\mathbf{p}^*),\ldots,\mathbf{x}_m(\mathbf{p}^*))$ maximizes social welfare.
\end{lemma}

\Cref{lem:SWDuality} tells us that the minimizer $\mathbf{p}^*$ of $f(\mathbf{p})=c^*(\mathbf{p})-\sum_{i=1}^{m}v_i^*(\mathbf{p})$ induces $\mathrm{SW}^*$, and thus it is natural to try to solve this dual optimization problem. However, $v_i$'s are unknown to us, and so are $v_i^*$'s and $f$. The second observation is that the revealed preference, $\mathbf{x}_i(\mathbf{p})$, actually gives a supergradient of $v_i^*$ at $\mathbf{p}$. Formally, given a concave continuous valuation $v:\mathcal{C}\to \mathbb{R}$, for any $\mathbf{x}\in \mathcal{C}$, by \Cref{conjgtSubgrad},
\begin{equation}\label{eq:RPSubgrad}
    \mathbf{x}\in\arg\max_{\mathbf{x}'\in \mathcal{C}}\del{v(\mathbf{x}')-\langle \mathbf{x}',\mathbf{p}\rangle}\iff \mathbf{x}\in\arg\min_{\mathbf{x}'\in \mathcal{C}}\del{\langle \mathbf{x}',\mathbf{p}\rangle-v(\mathbf{x}')}\iff\mathbf{x}\in\partial v^*(\mathbf{p}).
\end{equation}
Similarly, given a convex continuous cost $c:m \mathcal{C}\to \mathbb{R}_+$, for any $\mathbf{y}\in \mathbb{R}_+^n$, $\mathbf{y}\in\partial c^*(\mathbf{p})\iff \mathbf{y}\in\arg\max_{\mathbf{y}'}\del{\langle \mathbf{y}',\mathbf{p}\rangle-c(\mathbf{y}')}$. In other words, the subgradient of $c^*$ at $\mathbf{p}$ gives a bundle which maximizes the producer's profit, assuming everything produced can be sold.

As a result, we can run subgradient-based optimization algorithms to minimize $f$. ($c$ is known to the algorithm, and so is $c^*$; the computation of subgradients of $c^*$ is another problem, but does not require access to the consumer demand oracles.) Since $v_i$ is $\alpha$-strongly concave, by Lemma \ref{strongConvSmoothDual}, $v_i^*$ is $\nicefrac{1}{\alpha}$-strongly smooth and concave. However, there is no guarantee on $c^*$, and thus in general, $f$ is not strongly smooth. In this case the standard optimization algorithm is subgradient descent. However, for strongly smooth and convex functions, accelerated gradient descent converges much faster than subgradient descent. To invoke accelerated gradient descent, the \emph{smoothing technique} given in \citep{N05} is used. We minimize $f_{\mu}$ as given in \Cref{offlineSWMaxAlg} and tune the parameter $\mu$, which finally gives \Cref{offlineGuarantee}. Detailed proofs are given in \Cref{app_sec:offlineSW}.

\section{Online social welfare maximization}\label{sec:onlineSW}

\paragraph{Problem description.}

In online social welfare maximization, $m$ consumers come one by one and the producer/algorithm can post prices dynamically. Specifically, at step $i$, prices $\mathbf{p}_i$ are posted, and then consumer $i$ comes and makes a purchase $\mathbf{x}_i(\mathbf{p}_i)$. Then the algorithm updates $\mathbf{p}_i$ to $\mathbf{p}_{i+1}$, based on past information. The goal is to maximize the online social welfare $\sw(\mathbf{x}_1(\mathbf{p}_1),\ldots,\mathbf{x}_m(\mathbf{p}_m))=\sum_{i=1}^{m}v_i(\mathbf{x}_i(\mathbf{p}_i))-c(\sum_{i=1}^{m}\mathbf{x}_i(\mathbf{p}_i))$.

To model the randomness in the real world, it is usually assumed that valuations are sampled i.i.d. from some unknown distribution. Here we consider a slightly stronger model, called the random permutation model. In the random permutation model, an adversary chooses $m$ valuations $\tilde{v}_1,\dots,\tilde{v}_m$ in advance, which then come in a uniformly random order. Formally, let $\gamma=(\gamma_1,\dots,\gamma_m)$ be a random permutation of $(1,\dots,m)$, then at step $i$ the consumer with valuation $v_i=\tilde{v}_{\gamma_i}$ comes and makes a purchase, after $\mathbf{p}_i$ is posted. Note that in the random permutation model, the corresponding offline problem is fixed (with valuations $\tilde{v}_1,\ldots,\tilde{v}_m$). We still let $\sw^*=\max_{\mathbf{x}_1,\ldots,\mathbf{x}_m\in \mathcal{C}}\del{\sum_{i=1}^{m}\tilde{v}_i(\mathbf{x}_i)-c(\sum_{i=1}^{m}\mathbf{x}_i)}$, and let
\begin{equation*}
    (\tilde{\mathbf{x}}_1^*,\ldots,\tilde{\mathbf{x}}_m^*)=\arg\max_{\mathbf{x}_1,\ldots,\mathbf{x}_m\in \mathcal{C}}\del{\sum_{i=1}^{m}\tilde{v}_i(\mathbf{x}_i)-c\del{\sum_{i=1}^{m}\mathbf{x}_i}}.
\end{equation*}
Our goal is to show that the expected online social welfare $\mathbb{E}_{\gamma}[\sw(\mathbf{x}_1(\mathbf{p}_1),\ldots,\mathbf{x}_m(\mathbf{p}_m))]$ is close to $\sw^*$, where the expectation is taken over the random permutation $\gamma$.

Note that no more assumption is made; valuations are only required to be continuous.

\paragraph{Online convex optimization.}

The algorithm for online social welfare maximization invokes an online convex optimization (OCO) algorithm as a subroutine. In an OCO problem, there is a feasible domain $\mathcal{D}$ and $T$ steps. At step $t$, the OCO algorithm determines $\mathbf{x}_t\in \mathcal{D}$, and then a convex function $f_t:\mathcal{D}\to \mathbb{R}$ is chosen (potentially by an adversary) and a loss of $f_t(\mathbf{x}_t)$ is induced. Based on the past information (formally, $\mathbf{x}_1,\ldots,\mathbf{x}_t$ and $f_1,\ldots,f_t$), the algorithm updates $\mathbf{x}_t$ to $\mathbf{x}_{t+1}$, and tries to minimize the \emph{regret}
\begin{equation*}
    R(T)=\sum_{t=1}^{T}f_t(\mathbf{x}_t)-\min_{\mathbf{x}\in \mathcal{D}}\sum_{t=1}^{T}f_t(\mathbf{x}).
\end{equation*}
The regret of an OCO algorithm $\mathcal{A}$ is denoted by $R_{\mathcal{A}}(T)$.

The \emph{online (sub)gradient descent} algorithm performs the following update at step $t$:
\begin{equation*}
    \mathbf{x}_{t+1}=\Pi_{\mathcal{D}}[\mathbf{x}_t-\eta_tg_t(\mathbf{x}_t)],
\end{equation*}
where $\eta_t$ is the step size, $g_t(\mathbf{x}_t)\in\partial f_t(\mathbf{x}_t)$, and $\Pi_{\mathcal{D}}$ is the $\ell_2$ projection onto $\mathcal{D}$.
\begin{lemma}[\citep{H16} Theorem 3.1]\label{lem:ogd}
    Let $D=\max\{\|\mathbf{x}_1-\mathbf{x}_2\|_2|\mathbf{x}_1,\mathbf{x}_2\in \mathcal{D}\}$, $G=\max\{\|\partial f_t(\mathbf{x})\|_2|1\le t\le T,\mathbf{x}\in \mathcal{D}\}$, and $\eta_t=D/G\sqrt{T}$. Then
    \begin{equation*}
        R_{\mathrm{OGD}}(T)=\sum_{t=1}^{T}f_t(\mathbf{x}_t)-\min_{\mathbf{x}\in \mathcal{D}}\sum_{t=1}^{T}f_t(\mathbf{x})\le DG\sqrt{T}.
    \end{equation*}
\end{lemma}

\subsection{Algorithm and analysis}

\Cref{onlineSWMaxAlg} is proposed to solve the online social welfare maximization problem. The idea of \Cref{onlineSWMaxAlg} comes from \citep{AD15}.
\begin{algorithm}
    \caption{Online SW Maximization}
    \begin{algorithmic}
        \STATE $\mathcal{P}\leftarrow\{\mathbf{p}\ge \mathbf{0}|\|\mathbf{p}\|_2\le\lambda\}$.
        \STATE Give $\mathcal{P}$ to an OCO algorithm $\mathcal{A}$, and let $\mathbf{p}_1\in \mathcal{P}$ be the initial prices chosen by $\mathcal{A}$.
        \FOR {$i=1$ \TO $m$}
        \STATE Post prices $\mathbf{p}_i$.
        \STATE Observe $\mathbf{x}_i(\mathbf{p}_i)$, the choice of the buyer who shows up in the $i$-th step.
        \STATE Give $f_i(\mathbf{p})=\frac{1}{m}c^*(\mathbf{p})-\langle \mathbf{x}_i(\mathbf{p}_i),\mathbf{p}\rangle$ with domain $\mathcal{P}$ to $\mathcal{A}$, and observe an updated $\mathbf{p}_{i+1}$ from $\mathcal{A}$.
        \ENDFOR
    \end{algorithmic}
    \label{onlineSWMaxAlg}
\end{algorithm}
\begin{theorem}\label{onlineGuarantee}
    The expected social welfare of \Cref{onlineSWMaxAlg} with respect to a uniformly random permutation of continuous valuations, is within $R_{\mathcal{A}}(m)+2\lambda D_{\infty}\sqrt{nm}$ from the offline optimum social welfare, where $D_{\infty}=\max_{\mathbf{x},\mathbf{y}\in \mathcal{C}}\|\mathbf{x}-\mathbf{y}\|_{\infty}$ is the $\ell_{\infty}$ diameter of $\mathcal{C}$. Specifically, for online gradient descent, the difference is bounded by $4\lambda D_{\infty}\sqrt{nm}$.
\end{theorem}
\begin{proof}
For convenience, let $\mathbf{x}_i$ denote $\mathbf{x}_i(\mathbf{p}_i)$. By the regret bound of the OCO algorithm,
\begin{align}
    & \sum_{i=1}^{m}f_i(\mathbf{p}_i)-\min_{\mathbf{p}\ge \mathbf{0},\|\mathbf{p}\|_2\le\lambda}\sum_{i=1}^{m}f_i(\mathbf{p}) \nonumber \\
   =\  & \sum_{i=1}^{m}(f_i(\mathbf{p}_i)+v_i(\mathbf{x}_i))-\min_{\mathbf{p}\ge \mathbf{0},\|\mathbf{p}\|_2\le\lambda}\sum_{i=1}^{m}(f_i(\mathbf{p})+v_i(\mathbf{x}_i)) \label{onlineSWMaxRegret} \\
   \le\  & R_{\mathcal{A}}(m). \nonumber
\end{align}

First, we examine the second term of \cref{onlineSWMaxRegret}:
\begin{equation*}
    \begin{split}
        \min_{\mathbf{p}\ge \mathbf{0},\|\mathbf{p}\|_2\le\lambda}\sum_{i=1}^{m}(f_i(\mathbf{p})+v_i(\mathbf{x}_i)) & = \min_{\mathbf{p}\ge \mathbf{0},\|\mathbf{p}\|_2\le\lambda}\left(c^*(\mathbf{p})-\left\langle \sum_{i=1}^{m}\mathbf{x}_i,\mathbf{p}\right\rangle\right)+\sum_{i=1}^{m}v_i(\mathbf{x}_i) \\
         & =\sum_{i=1}^{m}v_i(\mathbf{x}_i)-c\left(\sum_{i=1}^{m}\mathbf{x}_i\right).
    \end{split}
\end{equation*}
The first equality comes from the definition of $f$, while the second inequality is due to the definition of $c^*$ and the monotonicity and Lipschitz continuity of $c$. Thus the second term of \cref{onlineSWMaxRegret} always equals the social welfare achieved by Algorithm \ref{onlineSWMaxAlg}. In the following we show that the first term of \cref{onlineSWMaxRegret} is within $O(\sqrt{mn})$ from the offline maximum social welfare $\sw^*$.

For a permutation $(\gamma_1,\ldots,\gamma_m)$ of $1,\ldots,m$, let $\Gamma_i$ denote $(\gamma_1,\ldots,\gamma_i)$. Note that $\mathbf{p}_i$ is determined by $\Gamma_{i-1}$ ($\Gamma_0=\emptyset$), $v_i$ is determined by $\gamma_i$, and $\mathbf{x}_i$ depends on $\mathbf{p}_i$ and $\gamma_i$. Fix $1\le i\le m$ and $\Gamma_{i-1}$, note that the revealed preference $\mathbf{x}_i$ maximizes the quasi-linear utility given $\mathbf{p}_i$, we have
\begin{align}
    \mathbb{E}_{\gamma_i}[f_i(\mathbf{p}_i)+v_i(\mathbf{x}_i)|\Gamma_{i-1}] & =\mathbb{E}_{\gamma_i}\left[\frac{1}{m}c^*(\mathbf{p}_i)-\langle \mathbf{x}_i,\mathbf{p}_i\rangle+v_i(\mathbf{x}_i)\middle|\Gamma_{i-1}\right] \nonumber \\
     & =\frac{1}{m}c^*(\mathbf{p}_i)+\mathbb{E}_{\gamma_i}[-\langle \mathbf{x}_i,\mathbf{p}_i\rangle+\tilde{v}_{\gamma_i}(\mathbf{x}_i)|\Gamma_{i-1}] \nonumber \\
     & \ge\frac{1}{m}c^*(\mathbf{p}_i)+\mathbb{E}_{\gamma_i}[-\langle \tilde{\mathbf{x}}_{\gamma_i}^*,\mathbf{p}_i\rangle+\tilde{v}_{\gamma_i}(\tilde{\mathbf{x}}_{\gamma_i}^*)|\Gamma_{i-1}] \nonumber \\
     & =\frac{1}{m}c^*(\mathbf{p}_i)-\langle \mathbb{E}_{\gamma_i}[\tilde{\mathbf{x}}_{\gamma_i}^*|\Gamma_{i-1}],\mathbf{p}_i\rangle+\mathbb{E}_{\gamma_i}[\tilde{v}_{\gamma_i}(\tilde{\mathbf{x}}_{\gamma_i}^*)|\Gamma_{i-1}]. \label{eq:onlineTmp1}
\end{align}
Consider the last term of \cref{eq:onlineTmp1} and take expectation with respect to $\Gamma_{i-1}$:
\begin{equation}\label{eq:onlineTmp2}
    \mathbb{E}_{\Gamma_{i-1}}\sbr{\mathbb{E}_{\gamma_i}[\tilde{v}_{\gamma_i}(\tilde{\mathbf{x}}_{\gamma_i}^*)|\Gamma_{i-1}]}=\frac{1}{m}\sum_{i=1}^{m}\tilde{v}_i(\tilde{\mathbf{x}}_i^*).
\end{equation}
Then consider the first two terms of \cref{eq:onlineTmp1}:
\begin{equation}\label{eq:onlineTmp3}
\begin{split}
    \frac{1}{m}c^*(\mathbf{p}_i)-\langle \mathbb{E}_{\gamma_i}[\tilde{\mathbf{x}}_{\gamma_i}^*|\Gamma_{i-1}],\mathbf{p}_i\rangle & =\frac{1}{m}c^*(\mathbf{p}_i)-\langle \mathbb{E}_{\gamma_i}[\tilde{\mathbf{x}}_{\gamma_i}^*],\mathbf{p}_i\rangle+\langle \mathbb{E}_{\gamma_i}[\tilde{\mathbf{x}}_{\gamma_i}^*],\mathbf{p}_i\rangle-\langle \mathbb{E}_{\gamma_i}[\tilde{\mathbf{x}}_{\gamma_i}^*|\Gamma_{i-1}],\mathbf{p}_i\rangle \\
     & \ge \frac{1}{m}c^*(\mathbf{p}_i)-\langle \mathbb{E}_{\gamma_i}[\tilde{\mathbf{x}}_{\gamma_i}^*],\mathbf{p}_i\rangle-\lambda\enVert{\mathbb{E}_{\gamma_i}[\tilde{\mathbf{x}}_{\gamma_i}^*]-\mathbb{E}_{\gamma_i}[\tilde{\mathbf{x}}_{\gamma_i}^*|\Gamma_{i-1}]}_2 \\
     & =\frac{1}{m}c^*(\mathbf{p}_i)-\frac{1}{m}\left\langle \sum_{i=1}^{m}\tilde{\mathbf{x}}_i^*,\mathbf{p}_i\right\rangle-\lambda\enVert{\mathbb{E}_{\gamma_i}[\tilde{\mathbf{x}}_{\gamma_i}^*]-\mathbb{E}_{\gamma_i}[\tilde{\mathbf{x}}_{\gamma_i}^*|\Gamma_{i-1}]}_2 \\
     & \ge-\frac{1}{m}c\del{\sum_{i=1}^{m}\tilde{\mathbf{x}}_i^*}-\lambda\enVert{\mathbb{E}_{\gamma_i}[\tilde{\mathbf{x}}_{\gamma_i}^*]-\mathbb{E}_{\gamma_i}[\tilde{\mathbf{x}}_{\gamma_i}^*|\Gamma_{i-1}]}_2.
\end{split}
\end{equation}
Here the first inequality is due to Cauchy-Schwarz inequality, while the second inequality is given by Fenchel-Young inequality.

\cref{eq:onlineTmp1}, \cref{eq:onlineTmp2} and \cref{eq:onlineTmp3} give us
\begin{equation}\label{eq:onlineTmp4}
    \mathbb{E}\left[\sum_{i=1}^{m}f_i(\mathbf{p}_i)+v_i(\mathbf{x}_i)\right]\ge \sum_{i=1}^{m}\tilde{v}_i(\tilde{\mathbf{x}}_i^*)-c\left(\sum_{i=1}^{m}\tilde{\mathbf{x}}_i^*\right)-\lambda \sum_{i=1}^{m}\mathbb{E}_{\Gamma_{i-1}}\left[\enVert{\mathbb{E}_{\gamma_i}[\tilde{\mathbf{x}}_{\gamma_i}^*]-\mathbb{E}_{\gamma_i}[\tilde{\mathbf{x}}_{\gamma_i}^*|\Gamma_{i-1}]}_2\right].
\end{equation}
Furthermore, \Cref{offlineOptError} shows that the last sum in \cref{eq:onlineTmp4} is bounded by $2D_{\infty}\sqrt{nm}$, and thus \Cref{onlineGuarantee} is proved for general OCO algorithms. Finally, to prove the bound for online gradient descent, it is enough to use step size $\eta_i=2\lambda/D\sqrt{m}$ (recall that $D$ is the $\ell_2$ diameter of $\mathcal{C}$) and invoke \Cref{lem:ogd}.
\end{proof}

As we can see from the proof of \Cref{onlineGuarantee}, it is enough to have continuous valuations. Furthermore, \Cref{onlineSWMaxAlg} still works if consumers only maximize their quasi-linear utilities approximately. Formally, if consumer $i$ finds a bundle $\mathbf{x}_i$ such that $v_i(\mathbf{x}_i)-\langle \mathbf{p}_i,\mathbf{x}_i\rangle\ge\max_{\mathbf{x}\in \mathcal{C}}\del{v_i(\mathbf{x})-\langle \mathbf{p}_i,\mathbf{x}\rangle}-\epsilon_i$, then an additive error of $\epsilon_i$ will be introduced in \cref{eq:onlineTmp1}. However, as long as the total error $\sum_{i=1}^{m}\epsilon_i$ is not large, the expected online social welfare of Algorithm \ref{onlineSWMaxAlg} will still be close to the offline optimum.

\section{Profit maximization for separable valuations and cost}\label{sec:profit}
Previously, social welfare maximization is solved by reducing to a convex optimization problem on the price space. However, profit maximization may be non-convex on both the bundle space and the price space.
\begin{example}\label{eg}
    Consider a market where there is only one consumer, one good, and zero cost. Suppose $v':[0,2]\to \mathbb{R}_+$ is continuous and strictly decreasing, with $v'(1)=2$, $v'(2)=1$, and $v'(x)x<2$ for any $x\ne1,2$. The integral of $v'$ gives a non-decreasing concave valuation $v$. It can be shown that the maximum profit is $2$, which is attained by price $2$ at quantity $1$ or price $1$ at quantity $2$. Thus the set of optimum prices and optimum bundles are both non-convex.
\end{example}

Here we present an algorithm of profit maximization and a nearly matching lower bound when all $v_i$'s and $c$ are separable. Formally, for every $\mathbf{x}\in \mathcal{C}$ and every $1\le i\le m$, $v_i(\mathbf{x})=\sum_{j=1}^{n}v_{ij}(x_j)$, and for every $\mathbf{y}\in m\mathcal{C}$, $c(\mathbf{y})=\sum_{j=1}^{n}c_j(y_j)$. Due to the separability assumption, we restate the assumptions on the feasible set, valuation functions and cost function:
\begin{itemize}
    \item $\mathcal{C}=[0,1]^n$.
    \item For every $1\le i\le m$, $1\le j\le n$, $v_{ij}$ is $\alpha$-strongly concave and $\lambda$-Lipschitz continuous.
    \item For every $1\le j\le n$, $c_j$ is $\lambda$-Lipschitz continuous.
\end{itemize}

The $i$-th consumer's consumption of good $j$ is completely determined by $p_j$ and is denoted by $x_{ij}(p_j)$. Furthermore, $x_j(p_j)=\sum_{i=1}^{m}x_{ij}(p_j)$. Our goal is thus to maximize $\mathrm{Profit}_j(p_j)=\sum_{j=1}^{n}x_j(p_j)p_j-c_j(x_j(p_j))$, for each $1\le j\le n$. Although we can set prices for different goods independently now, to keep consistency, we still consider posting new prices $\mathbf{p}\in \mathbb{R}_+^n$ as one query.
\begin{algorithm}
    \caption{Profit Maximization Algorithm for Separable Functions}
    \begin{algorithmic}
        \STATE $r\leftarrow\left\lceil \frac{mn\lambda(\lambda+\alpha)}{\alpha\epsilon}\right\rceil$.
        \STATE $\tilde{\mathbf{p}}=(0,0,\ldots,0)$.
        \FOR {$t=1$ \TO $r$}
            \STATE Post prices $\mathbf{p}_t=\left(\frac{t\alpha\epsilon}{mn(\lambda+\alpha)},\ldots,\frac{t\alpha\epsilon}{mn(\lambda+\alpha)}\right)$.
            \FOR {$j=1$ \TO $n$}
                \IF {$\mathrm{Profit}_j(p_{t,j})>\mathrm{Profit}_j(\tilde{p}_j)$}
                    \STATE {$\tilde{p}_j=p_{t,j}$}
                \ENDIF
            \ENDFOR
        \ENDFOR
        \STATE Output $\tilde{\mathbf{p}}$.
    \end{algorithmic}
    \label{profitMaxAlg}
\end{algorithm}

\begin{theorem}\label{thm:profitUpper}
    The profit given by Algorithm \ref{profitMaxAlg} is no less than the optimum profit minus $\epsilon$. The number of queries is $\lceil mn\lambda(\lambda+\alpha)/\alpha\epsilon\rceil$.
\end{theorem}
\begin{proof}
    Fix $1\le j\le n$. Let $p_j^*$ denote the profit-maximizing price for good $j$. Suppose $\hat{p}_j=\frac{z\alpha\epsilon}{mn(\lambda+\alpha)}\le p_j^*\le \frac{(z+1)\alpha\epsilon}{mn(\lambda+\alpha)}$. By the definition of strong smoothness and Lemma \ref{strongConvSmoothDual}, we have
    \begin{equation*}
        \begin{split}
            x_j(p_j^*)p_j^*-c_j(x_j(p_j^*)) & \le x_j(\hat{p}_j)\del{\hat{p}_j+\frac{\alpha\epsilon}{mn(\lambda+\alpha)}}-c_j(x_j(\hat{p}_j))+\lambda |x_j(p_j^*)-x_j(\hat{p}_j)| \\
             & =x_j(\hat{p}_j)\del{\hat{p}_j+\frac{\alpha\epsilon}{mn(\lambda+\alpha)}}-c_j(x_j(\hat{p}_j))+\lambda \left|\sum_{i=1}^{m}\del{(v_{ij}^*)'(p_j^*)-(v_{ij}^*)'(\hat{p}_j)}\right| \\
             & \le x_j(\hat{p}_j)\del{\hat{p}_j+\frac{\alpha\epsilon}{mn(\lambda+\alpha)}}-c_j(x_j(\hat{p}_j))+\lambda \frac{m}{\alpha}\frac{\alpha\epsilon}{mn(\lambda+\alpha)} \\
             & =x_j(\hat{p}_j)\hat{p}_j-c_j(x_j(\hat{p}_j))+x_j(\hat{p}_j)\frac{\alpha\epsilon}{mn(\lambda+\alpha)}+ \frac{\lambda\epsilon}{n(\lambda+\alpha)} \\
             & \le x_j(\hat{p}_j)\hat{p}_j-c_j(x_j(\hat{p}_j))+\frac{\alpha\epsilon}{n(\lambda+\alpha)}+ \frac{\lambda\epsilon}{n(\lambda+\alpha)} \\
             & =x_j(\hat{p}_j)\hat{p}_j-c_j(x_j(\hat{p}_j))+\frac{\epsilon}{n}.
        \end{split}
    \end{equation*}
\end{proof}
\begin{remark}
    Note that if the cost is $0$ and thus revenue maximization i considered, then we can set $r=\lceil mn\lambda/\epsilon\rceil$ in \Cref{profitMaxAlg}, and it is enough to assume concave valuations.
\end{remark}

\Cref{thm:profitLower} shows that the dependency on $n$ and $\epsilon$ cannot be improved, even for revenue maximization. The proof is given in \Cref{app_sec:profit}.
\begin{theorem}\label{thm:profitLower}
    The revenue maximization problem needs $\Omega(n/\epsilon)$ queries to get an additive error $\epsilon$, even if the valuations are separable, concave, non-decreasing, and Lipschitz continuous.
\end{theorem}

\section{Conclusion and open problems}

In this paper, we study social welfare and profit maximization with only revealed preferences. The social welfare maximization problem can be solved by reducing to a convex dual optimization problem in both the offline and online case, while profit maximization is essentially non-convex, for which we give nearly matching upper and lower bounds on the query complexity when valuations and cost are separable.

While social welfare maximization is interesting and important, it is still more reasonable for a producer to maximize profit. However, as shown by \Cref{eg}, this problem is in general non-convex. While we give an algorithm for the separable case, it is a very interesting open problem to design algorithms for profit maximization in a more general setting or show some hardness result.

\bibliography{paper}

\begin{thebibliography}{37}
\providecommand{\natexlab}[1]{#1}
\providecommand{\url}[1]{\texttt{#1}}
\expandafter\ifx\csname urlstyle\endcsname\relax
  \providecommand{\doi}[1]{doi: #1}\else
  \providecommand{\doi}{doi: \begingroup \urlstyle{rm}\Url}\fi

\bibitem[Afriat(1967)]{Afriat}
S.~N.. Afriat.
\newblock The construction of utility functions from expenditure data.
\newblock \emph{International Economic Review}, 1967.

\bibitem[Agrawal and Devanur(2015)]{AD15}
Shipra Agrawal and Nikhil~R Devanur.
\newblock Fast algorithms for online stochastic convex programming.
\newblock In \emph{Proceedings of the Twenty-Sixth Annual ACM-SIAM Symposium on
  Discrete Algorithms}, pages 1405--1424. SIAM, 2015.

\bibitem[Allen-Zhu and Orecchia(2014)]{AO14}
Zeyuan Allen-Zhu and Lorenzo Orecchia.
\newblock Linear coupling: An ultimate unification of gradient and mirror
  descent.
\newblock \emph{arXiv preprint arXiv:1407.1537}, 2014.

\bibitem[Amin et~al.(2015)Amin, Cummings, Dworkin, Kearns, and Roth]{ACDKR15}
Kareem Amin, Rachel Cummings, Lili Dworkin, Michael Kearns, and Aaron Roth.
\newblock Online learning and profit maximization from revealed preferences.
\newblock In \emph{AAAI}, pages 770--776, 2015.

\bibitem[Babaioff et~al.(2015)Babaioff, Dughmi, Kleinberg, and
  Slivkins]{BDKS15}
Moshe Babaioff, Shaddin Dughmi, Robert Kleinberg, and Aleksandrs Slivkins.
\newblock Dynamic pricing with limited supply.
\newblock \emph{ACM Transactions on Economics and Computation}, 3\penalty0
  (1):\penalty0 4, 2015.

\bibitem[Balcan et~al.(2014)Balcan, Daniely, Mehta, Urner, and
  Vazirani]{BDMUV14}
Maria-Florina Balcan, Amit Daniely, Ruta Mehta, Ruth Urner, and Vijay~V
  Vazirani.
\newblock Learning economic parameters from revealed preferences.
\newblock In \emph{International Conference on Web and Internet Economics},
  pages 338--353. Springer, 2014.

\bibitem[Beigman and Vohra(2006)]{BV06}
Eyal Beigman and Rakesh Vohra.
\newblock Learning from revealed preference.
\newblock In \emph{Proceedings of the 7th ACM Conference on Electronic
  Commerce}, pages 36--42. ACM, 2006.

\bibitem[Besbes and Zeevi(2009)]{BZ09}
Omar Besbes and Assaf Zeevi.
\newblock Dynamic pricing without knowing the demand function: Risk bounds and
  near-optimal algorithms.
\newblock \emph{Operations Research}, 57\penalty0 (6):\penalty0 1407--1420,
  2009.

\bibitem[Besbes and Zeevi(2012)]{BZ12}
Omar Besbes and Assaf Zeevi.
\newblock Blind network revenue management.
\newblock \emph{Operations research}, 60\penalty0 (6):\penalty0 1537--1550,
  2012.

\bibitem[Blum et~al.(2015)Blum, Mansour, and Morgenstern]{BMM15}
Avrim Blum, Yishay Mansour, and Jamie Morgenstern.
\newblock Learning what's going on: reconstructing preferences and priorities
  from opaque transactions.
\newblock In \emph{Proceedings of the Sixteenth ACM Conference on Economics and
  Computation}, pages 601--618. ACM, 2015.

\bibitem[Broder and Rusmevichientong(2012)]{BR12}
Josef Broder and Paat Rusmevichientong.
\newblock Dynamic pricing under a general parametric choice model.
\newblock \emph{Operations Research}, 60\penalty0 (4):\penalty0 965--980, 2012.

\bibitem[den Boer(2015)]{dB15}
Arnoud~V den Boer.
\newblock Dynamic pricing and learning: historical origins, current research,
  and new directions.
\newblock \emph{Surveys in operations research and management science},
  20\penalty0 (1):\penalty0 1--18, 2015.

\bibitem[den Boer and Zwart(2013)]{dBZ13}
Arnoud~V den Boer and Bert Zwart.
\newblock Simultaneously learning and optimizing using controlled variance
  pricing.
\newblock \emph{Management science}, 60\penalty0 (3):\penalty0 770--783, 2013.

\bibitem[Devanur and Hayes(2009)]{DH09}
Nikhil~R Devanur and Thomas~P Hayes.
\newblock The adwords problem: online keyword matching with budgeted bidders
  under random permutations.
\newblock In \emph{Proceedings of the 10th ACM conference on Electronic
  commerce}, pages 71--78. ACM, 2009.

\bibitem[Diewert(1973)]{die}
E.~Diewert.
\newblock Afriat and revealed preference theory.
\newblock \emph{Review of Economic Studies}, 40:\penalty0 419--426, 1973.

\bibitem[Dong et~al.(2018)Dong, Roth, Schutzman, Waggoner, and Wu]{DRSWW18}
Jinshuo Dong, Aaron Roth, Zachary Schutzman, Bo~Waggoner, and Zhiwei~Steven Wu.
\newblock Strategic classification from revealed preferences.
\newblock In \emph{Proceedings of the 2018 ACM Conference on Economics and
  Computation}, pages 55--70. ACM, 2018.

\bibitem[Goel and Mehta(2008)]{GM08}
Gagan Goel and Aranyak Mehta.
\newblock Online budgeted matching in random input models with applications to
  adwords.
\newblock In \emph{Proceedings of the nineteenth annual ACM-SIAM symposium on
  Discrete algorithms}, pages 982--991. Society for Industrial and Applied
  Mathematics, 2008.

\bibitem[Hazan et~al.(2016)]{H16}
Elad Hazan et~al.
\newblock Introduction to online convex optimization.
\newblock \emph{Foundations and Trends{\textregistered} in Optimization},
  2\penalty0 (3-4):\penalty0 157--325, 2016.

\bibitem[Hiriart-Urruty and Lemar{\'e}chal(2012)]{HL12}
Jean-Baptiste Hiriart-Urruty and Claude Lemar{\'e}chal.
\newblock \emph{Fundamentals of convex analysis}.
\newblock Springer Science \& Business Media, 2012.

\bibitem[Houthakker(1950)]{HSH}
H.~S. Houthakker.
\newblock Revealed preference and the utility function.
\newblock \emph{Economica}, 17:\penalty0 159--174, 1950.

\bibitem[Keskin and Zeevi(2014)]{KZ14}
N~Bora Keskin and Assaf Zeevi.
\newblock Dynamic pricing with an unknown demand model: Asymptotically optimal
  semi-myopic policies.
\newblock \emph{Operations Research}, 62\penalty0 (5):\penalty0 1142--1167,
  2014.

\bibitem[Mas-Colell(1977)]{MC-RP1}
Andreu Mas-Colell.
\newblock The recoverability of consumers' preferences from market demand.
\newblock \emph{Econometrica}, 45\penalty0 (6):\penalty0 1409--1430, 1977.

\bibitem[Mas-Colell(1978)]{MC-RP2}
Andreu Mas-Colell.
\newblock On revealed preference analysis.
\newblock \emph{The Review of Economic Studies}, 45\penalty0 (1):\penalty0
  121--131, 1978.

\bibitem[Nesterov(2005)]{N05}
Yu~Nesterov.
\newblock Smooth minimization of non-smooth functions.
\newblock \emph{Mathematical programming}, 103\penalty0 (1):\penalty0 127--152,
  2005.

\bibitem[Nesterov(1983)]{N83}
Yurii Nesterov.
\newblock A method of solving a convex programming problem with convergence
  rate $o(1/k^2)$.
\newblock \emph{Soviet Math. Dokl.}, 27\penalty0 (2):\penalty0 372--376, 1983.

\bibitem[Nesterov(2013)]{N13}
Yurii Nesterov.
\newblock \emph{Introductory lectures on convex optimization: A basic course}.
\newblock Springer Science \& Business Media, 2013.

\bibitem[Richter(1966)]{Rich}
M.~Richter.
\newblock Revealed preference theory.
\newblock \emph{Econometrica}, 34\penalty0 (3):\penalty0 635--645, 1966.

\bibitem[Rockafellar(1970)]{ROC}
R.~Tyrrell Rockafellar.
\newblock \emph{Convex Analysis}.
\newblock Princeton University Press, 1970.

\bibitem[Roth et~al.(2016)Roth, Ullman, and Wu]{RUW16}
Aaron Roth, Jonathan Ullman, and Zhiwei~Steven Wu.
\newblock Watch and learn: Optimizing from revealed preferences feedback.
\newblock In \emph{Proceedings of the forty-eighth annual ACM symposium on
  Theory of Computing}, pages 949--962. ACM, 2016.

\bibitem[Roth et~al.(2017)Roth, Slivkins, Ullman, and Wu]{RSUW17}
Aaron Roth, Aleksandrs Slivkins, Jonathan Ullman, and Zhiwei~Steven Wu.
\newblock Multidimensional dynamic pricing for welfare maximization.
\newblock In \emph{Proceedings of the 2017 ACM Conference on Economics and
  Computation}, pages 519--536. ACM, 2017.

\bibitem[Samuelson(1938)]{S38}
Paul~A Samuelson.
\newblock A note on the pure theory of consumer's behaviour.
\newblock \emph{Economica}, 5\penalty0 (17):\penalty0 61--71, 1938.

\bibitem[Sion et~al.(1958)]{sion}
Maurice Sion et~al.
\newblock On general minimax theorems.
\newblock \emph{Pacific Journal of mathematics}, 8\penalty0 (1):\penalty0
  171--176, 1958.

\bibitem[Tseng(2008)]{T08}
Paul Tseng.
\newblock On accelerated proximal gradient methods for convex-concave
  optimization.
\newblock \url{http://www.mit.edu/~dimitrib/PTseng/papers/apgm.pdf}, 2008.

\bibitem[Uzawa(1960)]{uzawa}
H.~Uzawa.
\newblock Preference and rational choice in the theory of consumption.
\newblock \emph{Mathematical Models in Social Science, eds. K. J. Arrow, S.
  Karlin, and P. Suppes}, 1960.

\bibitem[Varian(2005)]{varian05}
Hal~R. Varian.
\newblock Revealed preference.
\newblock In \emph{Samuelsonian Economics and the 21st Century. M. Szenberg, L.
  Ramrattand, and A. A. Gottesman, editors}, pages 99--115, 2005.

\bibitem[Wang et~al.(2014)Wang, Deng, and Ye]{WDY14}
Zizhuo Wang, Shiming Deng, and Yinyu Ye.
\newblock Close the gaps: A learning-while-doing algorithm for single-product
  revenue management problems.
\newblock \emph{Operations Research}, 62\penalty0 (2):\penalty0 318--331, 2014.

\bibitem[Zadimoghaddam and Roth(2012)]{ZR12}
Morteza Zadimoghaddam and Aaron Roth.
\newblock Efficiently learning from revealed preference.
\newblock In \emph{WINE}, pages 114--127. Springer, 2012.

\end{thebibliography}
\bibliographystyle{plainnat}

\appendix

\section{Omitted proofs from \Cref{sec:offlineSW}}\label{app_sec:offlineSW}

\begin{proof}[Proof of \Cref{lem:SWDuality}]
    Maximizing social welfare is equivalent to solving the following problem
    \begin{equation}\label{socWelfMaxUnboundedProg}
        \begin{array}{rl}
            \displaystyle\max_{\substack{\mathbf{x}_1,\ldots,\mathbf{x}_m\in \mathcal{C}\\\mathbf{y}\in m\mathcal{C}}} & \sum_{i=1}^{m}v_i(\mathbf{x}_i)-c(\mathbf{y}) \\
            \mathrm{s.t.} & \sum_{i=1}^{m}\mathbf{x}_i=\mathbf{y}.
        \end{array}
    \end{equation}
    The Lagrangian is $L(\mathbf{x}_1,\ldots,\mathbf{x}_m,\mathbf{y},\mathbf{p})=\sum_{i=1}^{m}v_i(\mathbf{x}_i)-c(\mathbf{y})+\langle \mathbf{p},\mathbf{y}-\sum_{i=1}^{m}\mathbf{x}_i\rangle$. \cref{socWelfMaxUnboundedProg} equals
    \begin{equation}
        \begin{split}
             & \max_{\mathbf{x}_1,\ldots,\mathbf{x}_m\in \mathcal{C},\mathbf{y}\in m\mathcal{C}}\min_{\mathbf{p}\in \mathbb{R}^n}L(\mathbf{x}_1,\ldots,\mathbf{x}_m,\mathbf{y},\mathbf{p}) \\
            = & \min_{\mathbf{p}\in \mathbb{R}^n}\max_{\mathbf{x}_1,\ldots,\mathbf{x}_m\in \mathcal{C},\mathbf{y}\in m\mathcal{C}}L(\mathbf{x}_1,\ldots,\mathbf{x}_m,\mathbf{y},\mathbf{p}) \\
            = & \min_{\mathbf{p}\in \mathbb{R}^n}\del{c^*(\mathbf{p})-\sum_{i=1}^{m}v_i^*(\mathbf{p})}.
        \end{split}
    \end{equation}
    Here the second line is due to Slater's condition, and the third line is due to the definition of convex conjugate and concave conjugate.

    Due to the continuity of $v_i$'s and $c$ and the compactness of $\mathcal{C}$, the optimal primal solution $(\mathbf{x}_1^*,\ldots,\mathbf{x}_m^*,\mathbf{y}^*)$ exists. By Slater's condition, the dual optimal solution $\mathbf{p}^*$ also exists. By the minimax property, we know that $\mathbf{y}^*=\sum_{i=1}^{m}\mathbf{x}_i^*$, and $\mathbf{x}_i^*$ maximizes $v_i(\mathbf{x})-\langle \mathbf{x},\mathbf{p}^*\rangle$, $1\le i\le m$.

    For the second part, due to the monotonicity of $c$, maximizing social welfare is equivalent to
    \begin{equation}\label{socWelfMaxProg}
        \begin{array}{rl}
            \displaystyle\max_{\substack{\mathbf{x}_1,\ldots,\mathbf{x}_m\in \mathcal{C}\\\mathbf{y}\in m \mathcal{C}}} & \sum_{i=1}^{m}v_i(\mathbf{x}_i)-c(\mathbf{y}) \\
            \mathrm{s.t.} & \sum_{i=1}^{m}\mathbf{x}_i\le \mathbf{y}.
        \end{array}
    \end{equation}
    The Lagrangian is still $L(\mathbf{x}_1,\ldots,\mathbf{x}_m,\mathbf{y},\mathbf{p})=\sum_{i=1}^{m}v_i(\mathbf{x}_i)-c(\mathbf{y})+\langle \mathbf{p},\mathbf{y}-\sum_{i=1}^{m}\mathbf{x}_i\rangle$. \cref{socWelfMaxProg} equals
    \begin{equation}
        \begin{split}
             & \max_{\mathbf{x}_1,\ldots,\mathbf{x}_m\in \mathcal{C},\mathbf{y}\in m \mathcal{C}}\min_{\mathbf{p}\ge \mathbf{0},\|\mathbf{p}\|_2\le\lambda}L(\mathbf{x}_1,\ldots,\mathbf{x}_m,\mathbf{y},\mathbf{p}) \\
            = & \min_{\mathbf{p}\ge \mathbf{0},\|\mathbf{p}\|_2\le\lambda}\max_{\mathbf{x}_1,\ldots,\mathbf{x}_m\in \mathcal{C},\mathbf{y}\in m \mathcal{C}}L(\mathbf{x}_1,\ldots,\mathbf{x}_m,\mathbf{y},\mathbf{p}) \\
            = & \min_{\mathbf{p}\ge \mathbf{0},\|\mathbf{p}\|_2\le\lambda}\del{c^*(\mathbf{p})-\sum_{i=1}^{m}v_i^*(\mathbf{p})}.
        \end{split}
    \end{equation}
    Here the first line is due to the Lipschitz continuity of $c$, the second line is due to Sion's theorem \citep{sion}.
\end{proof}

Now we come to the proof of \Cref{offlineGuarantee}. We first consider the special case where $c$ is $\mu$-strongly convex. Then $c^*$ is $\nicefrac{1}{\mu}$-strongly smooth and convex. Since $v_i$ is $\alpha$-strongly concave, by Lemma \ref{strongConvSmoothDual}, $v_i^*$ is $\nicefrac{1}{\alpha}$-strongly smooth and concave. Therefore $f$ is $(\nicefrac{1}{\mu}+\nicefrac{m}{\alpha})$-strongly smooth and convex. Invoke the AGM algorithm introduced in Lemma \ref{AGM} directly on $f$, and let $\mathbf{p}^T$ denote the output of the AGM algorithm after $T$ queries.
\begin{lemma}\label{offlineGuaranteeStrongConv}
    Let $\mathbf{p}^0=\mathbf{0}$, then for $T\ge1$, after $T$ queries to the aggregate demand oracle, the social welfare induced by $\mathbf{p}^T$ outputted by the AGM algorithm is within $4\lambda^2(\nicefrac{1}{\mu}+\nicefrac{m}{\alpha})/T$ from the maximum social welfare.
\end{lemma}
\begin{proof}
    Theorem 2.1.5 in \citep{N13} tells us that, $f:\mathbb{R}^n\to \mathbb{R}$ is $\beta$-strongly smooth and convex if and only if for any $\mathbf{x},\mathbf{y}\in \mathbb{R}^n$,
    \begin{equation}\label{strongSmoothEquiv}
        f(\mathbf{y})-f(\mathbf{x})\ge \langle\nabla f(\mathbf{x}),\mathbf{y}-\mathbf{x}\rangle+\frac{1}{2\beta}\|\nabla f(\mathbf{y})-\nabla f(\mathbf{x})\|_2^2.
    \end{equation}

    Let $\mathbf{p}^*\in\arg\min_{\mathbf{p}\in \mathcal{P}}f(\mathbf{p})$. By \Cref{lem:SWDuality}, we know that $\nabla f(\mathbf{p}^*)=\mathbf{0}$. Let $\mathbf{y}^T=\nabla c^*(\mathbf{p}^T)$, $\mathbf{x}_{i}^T=\nabla v_i^*(\mathbf{p}^T)$, and $\mathbf{x}_i^*=\nabla v_i^*(\mathbf{p}^*)$. Then by Lemma \ref{AGM} and \cref{strongSmoothEquiv}, we know that
    \begin{equation*}
        \left\|\mathbf{y}^T-\sum_{i=1}^{m}\mathbf{x}_{i}^T\right\|_2=\|\nabla f(\mathbf{p}^T)\|_2\le\sqrt{2\left(\frac{1}{\mu}+\frac{m}{\alpha}\right)(f(\mathbf{p}^T)-f(\mathbf{p}^*))}\le \frac{2\lambda(\frac{1}{\mu}+\frac{m}{\alpha})}{T}.
    \end{equation*}
    Then we have
    \begin{equation*}
        \begin{split}
             & \sum_{i=1}^{m}v_i(\mathbf{x}_{i}^T)-c\left(\sum_{i=1}^{m}\mathbf{x}_{i}^T\right) \\
            \ge\  & \sum_{i=1}^{m}v_i(\mathbf{x}_{i}^T)-c(\mathbf{y}^T)-\lambda\left\|\mathbf{y}^T-\sum_{i=1}^{m}\mathbf{x}_{i}^T\right\|_2 \\
            =\  & \sum_{i=1}^{m}(\langle \mathbf{x}_{i}^T,\mathbf{p}^T\rangle-v_i^*(\mathbf{p}^T))-\langle \mathbf{y}^T,\mathbf{p}^T\rangle+c^*(\mathbf{p}^T)-\lambda\left\|\mathbf{y}^T-\sum_{i=1}^{m}\mathbf{x}_{i}^T\right\|_2 \\
            \ge\  & c^*(\mathbf{p}^*)-\sum_{i=1}^{m}v_i^*(\mathbf{p}^*)-2\lambda\left\|\mathbf{y}^T-\sum_{i=1}^{m}\mathbf{x}_{i}^T\right\|_2. \\
        \end{split}
    \end{equation*}
    Here the first inequality comes from the Lipschitz continuity of $c$. As given by \Cref{lem:SWDuality}, $c^*(\mathbf{p}^*)-\sum_{i=1}^{m}v_i^*(\mathbf{p}^*)=\mathrm{SW}^*$. Thus the additive error between the social welfare achieved by $\mathbf{p}_T$ and the optimum social welfare is at most
    \begin{equation*}
        2\lambda\left\|\mathbf{y}^T-\sum_{i=1}^{m}\mathbf{x}_{i}^T\right\|_2\le \frac{4\lambda^2(\frac{1}{\mu}+\frac{m}{\alpha})}{T}.
    \end{equation*}
\end{proof}

For the general case, we make use of the \emph{smoothing technique}, given in \citep{N05}: Let $c_{\mu}(\mathbf{y})=c(\mathbf{y})+\frac{\mu}{2}\|\mathbf{y}\|_2^2$. Now $c_{\mu}$ is $\mu$-strongly convex, $(\lambda+\mu mD)$-Lipschitz continuous, and $c\le c_{\mu}\le c+\frac{\mu}{2}m^2D^2$. We can then just use $c_{\mu}$ as the cost function, and try to tune $\mu$.

\begin{proof}[Proof of \Cref{offlineGuarantee}]
    Let $\mathbf{p}^*\in\arg\min_{\mathbf{p}\in \mathcal{P}}f(\mathbf{p})$, $\mathbf{p}_{\mu}^*\in\arg\min_{\mathbf{p}\in \mathcal{P}}f_{\mu}(\mathbf{p})$, and $\mathbf{x}_{\mu,i}^{T}=\nabla v_i^*(\mathbf{p}_{\mu}^{T})$, $\mathbf{x}_{\mu,i}^*=\nabla v_i^*(\mathbf{p}_{\mu}^*)$, $\mathbf{x}_i^*=\nabla v_i^*(\mathbf{p}^*)$. We have
    \begin{equation}\label{offlineGuaranteeHelper1}
        \sum_{i=1}^{m}v_i(\mathbf{x}_{\mu,i}^{T})-c\left(\sum_{i=1}^{m}\mathbf{x}_{\mu,i}^{T}\right)\ge \sum_{i=1}^{m}v_i(\mathbf{x}_{\mu,i}^{T})-c_{\mu}\left(\sum_{i=1}^{m}\mathbf{x}_{\mu,i}^{T}\right),
    \end{equation}
    and
    \begin{equation}\label{offlineGuaranteeHelper2}
        \sum_{i=1}^{m}v_i(\mathbf{x}_{\mu,i}^*)-c_{\mu}\left(\sum_{i=1}^{m}\mathbf{x}_{\mu,i}^*\right)\ge \sum_{i=1}^{m}v_i(\mathbf{x}_i^*)-c_{\mu}\left(\sum_{i=1}^{m}\mathbf{x}_i^*\right)\ge \sum_{i=1}^{m}v_i(\mathbf{x}_i^*)-c\left(\sum_{i=1}^{m}\mathbf{x}_i^*\right)-\frac{\mu}{2}m^2D^2.
    \end{equation}
    Combine \cref{offlineGuaranteeHelper1}, \cref{offlineGuaranteeHelper2}, and the proof of Lemma \ref{offlineGuaranteeStrongConv}, we know that
    \begin{equation}
        \left(\sum_{i=1}^{m}v_i(\mathbf{x}_{\mu,i}^{T})-c\left(\sum_{i=1}^{m}\mathbf{x}_{\mu,i}^{T}\right)\right)-\left(\sum_{i=1}^{m}v_i(\mathbf{x}_i^*)-c\left(\sum_{i=1}^{m}\mathbf{x}_i^*\right)\right)\le \frac{\mu}{2}m^2D^2+ \frac{4(\lambda+\mu mD)^2(\frac{1}{\mu}+\frac{m}{\alpha})}{T}.
    \end{equation}
    The theorem is proved by setting $\mu=\frac{2\lambda}{mD\sqrt{T}}$.
\end{proof}

\section{Omitted proofs from \Cref{sec:onlineSW}}\label{app_sec:onlineSW}

\begin{lemma}\label{offlineOptError}
    For any $1\le i\le m$, we have
    \begin{equation*}
        \mathbb{E}_{\Gamma_{i-1}}\left[\enVert{\mathbb{E}_{\gamma_i}[\tilde{\mathbf{x}}_{\gamma_i}^*]-\mathbb{E}_{\gamma_i}[\tilde{\mathbf{x}}_{\gamma_i}^*|\Gamma_{i-1}]}_2\right]\le D_{\infty}\sqrt{\frac{n}{m-i+1}}.
    \end{equation*}
\end{lemma}
\begin{proof}
    Let $\mathcal{S}=\{s_1,\ldots,s_N\}$ denote a finite population of real numbers, and $X_1,\ldots,X_n$ ($1\le n\le N$) denote $n$ samples from $\mathcal{S}$ without replacement. Furthermore, let $\mu=\frac{1}{N}\sum_{i=1}^{N}s_i$ and $\sigma^2=\frac{1}{N}\sum_{i=1}^{N}(s_i-\mu)^2$. Then $\overline{X}=\frac{1}{n}\sum_{i=1}^{n}X_i$ has mean $\mu$ and variance $\frac{N-n}{N-1}\frac{\sigma^2}{n}$.

    Now come back the the proof of Lemma \ref{offlineOptError}. By Jensen's inequality, we have
    \begin{equation*}
        \mathbb{E}_{\Gamma_{i-1}}\left[\enVert{\mathbb{E}_{\gamma_i}[\tilde{\mathbf{x}}_{\gamma_i}^*]-\mathbb{E}_{\gamma_i}[\tilde{\mathbf{x}}_{\gamma_i}^*|\Gamma_{i-1}]}_2\right]\le\sqrt{\mathbb{E}_{\Gamma_{i-1}}\left[\enVert{\mathbb{E}_{\gamma_i}[\tilde{\mathbf{x}}_{\gamma_i}^*]-\mathbb{E}_{\gamma_i}[\tilde{\mathbf{x}}_{\gamma_i}^*|\Gamma_{i-1}]}_2^2\right]}.
    \end{equation*}
    Note that each coordinate of $\mathbb{E}_{\Gamma_{i-1}}\left[\|\mathbb{E}_{\gamma_i}[\tilde{\mathbf{x}}_{\gamma_i}^*]-\mathbb{E}_{\gamma_i}[\tilde{\mathbf{x}}_{\gamma_i}^*|\Gamma_{i-1}]\|_2^2\right]$ equals the variance of the average of $m-i+1$ without-replacement samples. Thus we further have
    \begin{equation*}
        \sqrt{\mathbb{E}_{\Gamma_{i-1}}\left[\enVert{\mathbb{E}_{\gamma_i}[\tilde{\mathbf{x}}_{\gamma_i}^*]-\mathbb{E}_{\gamma_i}[\tilde{\mathbf{x}}_{\gamma_i}^*|\Gamma_{i-1}]}_2^2\right]}\le\sqrt{n \frac{i-1}{m-1}\frac{D_{\infty}^2}{m-i+1}}\le D_{\infty}\sqrt{\frac{n}{m-i+1}}.
    \end{equation*}
\end{proof}

\section{Omitted proofs from \Cref{sec:profit}}\label{app_sec:profit}

\begin{proof}[Proof of \Cref{thm:profitLower}]
    Let us first consider the case where there is only one consumer and one good. Given $\lambda>1$, consider $\epsilon$ such that there exists an integer $q$ satisfying $(1+\epsilon)^q=\lambda$, and thus $q=\frac{\ln\lambda}{\ln(1+\epsilon)}\ge \frac{\ln\lambda}{\epsilon}$.

    Now we are going to define concave functions of the amount of good on $[0,1]$. It is enough to give a non-decreasing and integrable derivative. Let $v'(x)=\lambda$ on $[0,\frac{1}{\lambda}]$, and $\frac{1}{x}$ on $[\frac{1}{\lambda},1]$. One can verify that $v'$ is non-decreasing and integrable, and thus we can integrate it into a concave function $v$ (by shifting, we can ensure $v(0)=0$). The maximum of $v'(x)x$ on $[0,1]$ is $1$.

    We claim that the algorithm has to make at least $q$ queries to ensure an $\epsilon$ additive or multiplicative error. If it is not true, there must exist some integer $z\in[0,q-1]$ such that no $x\in \mathcal{I}=(\frac{1}{(1+\epsilon)^{z+1}},\frac{1}{(1+\epsilon)^z}]$ is considered. We can then set $\tilde{v}'(x)=v'(x)$ outside $\mathcal{I}$, $\tilde{v}'(x)=(1+\epsilon)^{z+1}$ on $(\frac{1}{(1+\epsilon)^{z+1}},\frac{1}{(1+\epsilon)^z})$, and $\tilde{v}'(\frac{1}{(1+\epsilon)^z})$ does not exist. $\tilde{v}'$ is still non-decreasing and integrable, but the optimum revenue is $1+\epsilon$ now, which is not detected by the algorithm.

    In other words, for $\lambda$-Lipschitz valuations, at least $\frac{\ln\lambda}{\epsilon}$ queries is required. W.l.o.g., suppose $\lambda=e$ and thus $\frac{1}{\epsilon}$ queries is needed. Now suppose there are $n$ goods, $T_j$ different prices are tested for the $j$-th good, and the profit we get from good $j$ is within $\epsilon_j$ from the maximum profit. Then $T_j\ge \frac{1}{\epsilon_j}$. We have
    \begin{equation*}
        \frac{\sum_{j=1}^{n}T_j}{n}\ge \frac{n}{\sum_{j=1}^{n}1/T_j}\ge \frac{n}{\sum_{j=1}^{n}\epsilon_j}=\frac{n}{\epsilon}.
    \end{equation*}
    In other words, $\sum_{j=1}^{n}T_j\ge \frac{n^2}{\epsilon}$. Since each query can set new prices for $n$ goods, we need $\Omega(\frac{n}{\epsilon})$ queries.
\end{proof}

\end{document}